\documentclass[twocolumn]{article}
\usepackage{epsfig,amsfonts}

\newcommand{\cj}[1]{\overline{#1}}
\newcommand{\mbf}[1]{\mbox{\boldmath \(#1\)}}
\newcommand{\fig}[1]{Fig.~\ref{fig:#1}}
\newcommand{\equ}[1]{Eq.~(\ref{eq:#1})}
\newcommand{\re}{\mathop{\mathrm{Re}}}
\newcommand{\im}{\mathop{\mathrm{Im}}}
\newcommand{\Cset}{\mathbb{C}}

\title{Quantum theory of computation and relativistic physics}

\author{Alexander Yu.\ Vlasov\thanks{
E-Mails: \underline{Alexander.Vlasov@pobox.spbu.ru},
 alex@protection.spb.su, vlasov@physics.harvard.edu} \\
197101, Mira Street 8, IRH\\
St.--Petersburg, Russia}

\date{Jan 1997 \\
      (Jul 1999)}

\begin{document}

 \textwidth	508bp	
 \oddsidemargin -20bp	
 \evensidemargin-20bp
 \columnsep=12bp	
 \normalmarginpar
 \marginparwidth 36bp	
 \marginparsep   4bp	


\begin{titlepage}

\maketitle

\medskip

\begin{minipage}[c]{420pt}

\centerline{\large\tt quant-ph/9701027}

\begin{verbatim}
Title: Quantum Theory of Computation and Relativistic Physics
Authors: Alexander Yu. Vlasov (FCR/IRH, St.-Petersburg, Russia)
Comments: 6 pages, LaTeX2e, 2 columns, 4 PostScript figures are included by
  epsfig.sty; based on poster in Proc. PhysComp '96 Workshop (BU, Boston MA,
  22-24 Nov 1996) pp. 332-333, a "no-go" result for bounded quantum networks;
  v3/4 -- typos corrected, minor changes.
Report-no: RQC-VAY11/v4

  In the e-print is discussed a few steps to introducing of "vocabulary" of
relativistic physics in quantum theory of information and computation (QTI&C).
The behavior of a few simple quantum systems those are used as models in
QTI&C is tested by usual relativistic tools (transformation properties of
wave vectors, etc.). Massless and charged massive particles with spin 1/2
are considered. Field theory is also discussed briefly.
\end{verbatim}
\end{minipage}

\rule{420pt}{.4pt}

\medskip

\begin{abstract}
In the paper are described some steps for merger between relativistic quantum
theory and theory of computation. The first step is consideration of
transformation of {\em qubit\/} state due to rotation of coordinate system. The
Lorentz transformation is considered after that. The some new properties of this
transformation change usual model of qubit. The system of q$^2$bit
seems more fundamental relativistic model. It is shown also that such
model as {\em electron\/} is really such q$^2$bit system and for modelling
of qubit is necessary to use massless particle like electron neutrino.

The quantum field theory (QFT) is briefly discussed further. The wave vectors
of interacted particle now described by some operator and it can produce some
multiparticle (`{\em nonlinear\/}') effects.
\end{abstract}

\medskip


\medskip


\thispagestyle{empty}
\end{titlepage}

\section{Introduction}\label{intro}

The paper describes some approaches to {\em relativistic quantum theory
of computation}. The main purpose of the work is to consider essentially new
properties of {\em quantum computers\/} \cite{benioff:erase,feynman:simul,%
feynman:comp,deutsch:turing,deutsch:gates} due to relativistic phenomena
rather than some small corrections to nonrelativistic formulae.

\medskip

At first, in relativistic theory it is necessary to consider a {\em qubit\/}
in different coordinate systems. In simplest case it may be 3D local
rotations and $SU(2)$ spinors.

For consideration of temporal coordinate it is necessary to use Lorentz
transformations and 4D spinors. The more correct approach include full
Poincare group and quantum field theory.

\section{Qubit}\label{qbt}

A quantum two-state system is often called {\em quantum bit\/} or {\em `qubit'\/}
\cite{schumacher:qubit,bennet:review}.
Let us consider a particle with spin $1/2$ as a model of the qubit. The quantum state
of the system is $\psi = c_0\left|0\right> + c_1\left|1\right> $,
where $c_0$ and $c_1$ are complex numbers and the norm of $\psi$ is:

\begin{equation}
       ||\psi ||^2 \equiv \psi^{*} \psi = |c_0|^2 + |c_1|^2 = 1,
       \quad c_0, c_1 \in \Cset
\label{eq:norm1}
\end{equation}

\begin{figure}[!htb]
\centering
 \begin{picture}(160,160)
 \small
 \put(0,0){\epsfig{file=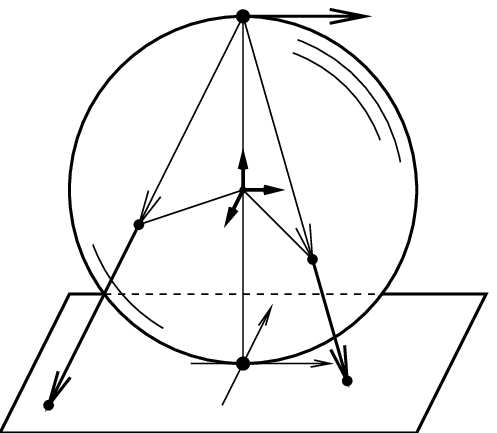}}
 \put(70,125){\bf False}
 \put(66,19){\makebox(0,0)[rt]{\bf True}}
 \put(136,20){\makebox(0,0)[l]{$\displaystyle 2 \times \frac{c_0}{c_1}$}}
 \put(110,120){\makebox(0,0)[l]{$\infty \ (c_1 = 0)$}}
 \put(80,34){\makebox(0,0)[l]{Im}}
 \put(90,17){\makebox(0,0)[t]{Re}}
 \put(68,78){\makebox(0,0)[br]{\it z}}
 \put(78,73){\makebox(0,0)[b]{\it x}}
 \put(62,60){\makebox(0,0)[b]{\it y}}
 \put(127,42){${\Cset}P$}
 \put(30,110){${\mathbf S}$}
 \end{picture}
\caption{Riemann sphere for qubit}\label{fig:qubit}
\end{figure}

A state of a qubit can be described as a superposition of two logical states of
usual bit ({\bf False}, {\bf True} or {\bf 0}, {\bf 1}) with complex
coefficients. The state of quantum system is described as a {\em ray\/} in
complex Hilbert space and for two-state system it can be considered as
complex projective space ${\Cset}P\sim{\Cset}\cup\{\infty\}$. Each ray
$(c_0,c_1)$ is presented by complex number $\zeta=c_0/c_1$. The $|0\rangle$
corresponds to $0$ and the $|1\rangle$ to $\infty$. There is correspondence
\cite{penrose:spinors} between the plane $\zeta$ and a sphere $\mathbf S$ due
to stereographic projection $\zeta = (x - \mathrm{i} y)/(1 - z)$
(see \fig{qubit}). Expressions for coordinate $(x,y,z)$ on the unit sphere are:
\begin{equation}
\begin{array}{rl}
 x  = & {\displaystyle\frac{2 \re\ \zeta}{|\zeta|^2 + 1}
    = \frac{c_0\cj{c}_1 + c_1\cj{c}_0}{c_0\cj{c}_0 + c_1\cj{c}_1}} \\
-y  = & {\displaystyle \frac{2 \im\ \zeta}{|\zeta|^2 + 1}
    = \frac{-\mathrm{i}(c_0\cj{c}_1 - c_1\cj{c}_0)}{c_0\cj{c}_0 + c_1\cj{c}_1}} \\
 z  = & {\displaystyle\frac{|\zeta|^2 - 1}{|\zeta|^2 + 1}
    = \frac{c_0\cj{c}_0 - c_1\cj{c}_1}{c_0\cj{c}_0 + c_1\cj{c}_1}}
\end{array}
\label{eq:coord3a}
\end{equation}
Due to the equation \equ{norm1} we can consider $(X,Y,Z)$ instead:
\begin{equation}
\begin{array}{l}
 X = c_0\cj{c}_1 + c_1\cj{c}_0\\
 Y = \mathrm{i}(c_0\cj{c}_1 - c_1\cj{c}_0) \\
 Z = c_0\cj{c}_0 - c_1\cj{c}_1
\end{array}
\label{eq:coord3}
\end{equation}
The $|0\rangle$ and $|1\rangle$ map to opposite poles of the sphere.

\subsection{Spatial rotation of coordinate system}

A transformation of the state due to a spatial rotation of coordinate
system is described by unitary matrix with determinant unity:

\begin{equation}
{
\psi' = \left(\matrix{a&b\cr
                        c&d\cr}\right) \psi, \quad
    \matrix{\overline{a} = d, \ \overline{c} = -b \cr
     a \, d - b \, c = |a|^2 + |b|^2 = 1}
}
\label{eq:transpin}
\end{equation}

This is the group of unitary $2 \times 2$ matrices, $SU(2)$. It corresponds
to principle, that transformation of the wave vector is described by some
representation of a group of coordinate transformation. The group $SU(2)$ is
representation of the group of spatial rotations $SO(3)$ in a space of 2D
complex vectors.

Due to 2--1 isomorphism $SU(2)$ and $SO(3)$, any rotation corresponds to
unitary matrix up to sign. We can see simple correspondence between any
1--gate and ``passive'' transformation, {\em i.e.\/} transition to other
coordinate system.

\medskip

The equations \equ{coord3} can be used for demonstration of relation
between $SO(3)$ and $SU(2)$. If we apply some unitary transformation \equ{transpin}
$U: (c_0,c_1) \to (c'_0,c'_1) $ then $(X,Y,Z) \to (X',Y',Z')$. Unitary
matrices do not change the norm \equ{norm1} and length of the vector:
\begin{equation}
 X^2 + Y^2 + Z^2 = (|c_0|^2+|c_1|^2)^2
\label{eq:norm3}
\end{equation}
Angles between vectors also do not change. Unitary transformations of a state
of the qubit correspond to rotations of the sphere (\fig{qubit}). Two matrices:
$\mathbf U$ and $\mathbf{-U}$ produce the same rotation due to \equ{coord3}.

\medskip

The transformations of a state of $n$--qubits due to spatial rotation can
be described by {\em unitary\/} $2^{2n}$ matrices.

\section{The relativistic consideration of a qubit}\label{qubit}

\subsection{Lorentz transformation}

For Lorentz transformation of coordinate system there is similar isomorphism
between the group $SO(3,1)$ and the group $SL(2,\Cset)$ of all complex
$2 \times 2$ matrices with determinant unity. The group $SL(2,\Cset)$ is
isomorphic with Lorentz group in the same way as the group $SU(2)$ with group
of 3D rotations \cite{penrose:spinors}. The group $SL(2,\Cset)$ is a
representation of Lorentz group $SO(3,1)$ in a space of 2D complex vectors.

On the other hand, we should not directly apply such representation of relativistic group
$SL(2,\Cset)$ to a qubit. Only the subgroup of unitary matrix saves the norm
\equ{norm1}. The expression \equ{norm1} in relativistic theory is not
invariant scalar, but temporal part of 4--vector. Simple relation between
transformations of coordinate system and unitary matrices is broken here.

Let us denote:
\begin{equation}
       T = ||\psi ||^2 \equiv \psi^{*} \psi = c_0\cj{c}_0 + c_1\cj{c}_1
\label{eq:coordT}
\end{equation}
We can write\footnote{
In the matrix notation $\psi^{*} \psi$ is scalar and $\psi \psi^*$ is
$2\times2$ matrix (with Dirac notation: $\langle\psi|\psi\rangle$ and
$|\psi\rangle\langle\psi|$ respectively).}
, using equations \equ{coord3}, \equ{coordT} :
\begin{equation}
\begin{array}{c}
\mathbf V \equiv \left(\matrix{T + Z & X - \mathrm{i} Y\cr
                        X + \mathrm{i} Y & T - Z\cr}\right) =
   2 \left(\matrix{c_0\cj{c}_0 & c_0\cj{c}_1 \cr
                  c_1\cj{c}_0 & c_1\cj{c}_1 \cr}\right) \\
\frac12 \mathbf V = \left(\matrix{c_0 \cr c_1 \cr}\right)
   \left(\matrix{\cj{c}_0 & \cj{c}_1 \cr}\right) = \psi \psi^* \\
\det \mathbf V = T^2 - X^2 - Y^2 - Z^2 = \\
   = 2 c_0\cj{c}_0 2 c_1\cj{c}_1 - 2 c_1\cj{c}_0 2 c_0\cj{c}_1 = 0

\end{array}
\label{eq:coord4}
\end{equation}

Linear transformations with determinant unity of a qubit correspond to Lorentz
transformation of the vector $(T,X,Y,Z)$:
\begin{equation}
\begin{array}{l}
\psi' = \mathbf A \psi; \quad \det \mathbf A = 1  \\
\mathbf V' = 2 \mathbf A \psi (\mathbf A \psi)^* =
             2 \mathbf A \psi \psi^* \mathbf A ^* = \mathbf{A V A}^* \\
\det \mathbf V' = T'^2 - X'^2 - Y'^2 - Z'^2 = \\
= \det \mathbf V  = T^2 - X^2 - Y^2 - Z^2
\end{array}
\label{eq:weyltr}
\end{equation}

\begin{figure}[!htb]
\centering
 \begin{picture}(140,190)
 \small
 \put(0,0){\epsfig{file=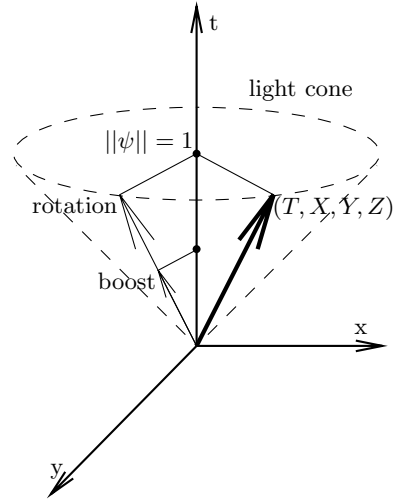}}
 \put(90,155){light cone}
 \put(75,180){t}
 \put(130,65){x}
 \put(15,12){y}
 \put(99,117){\makebox(0,0)[tl]{$(T,X,Y,Z)$}}
 \put(40.5,116.5){\makebox(0,0)[tr]{rotation}}
 \put(55,88){\makebox(0,0)[tr]{boost}}
 \put(69,133){\makebox(0,0)[br]{$||\psi||=1$}}
 \end{picture}
\caption{Null vector $(T,X,Y,Z)$}\label{fig:isotr}
\end{figure}

Only if the matrix $\mathbf A$ is unitary, $\mathbf{A V A}^* = \mathbf{A V A}^{-1}$
and {\em Trace \bf V} {\em i.e.\/} the norm \equ{coordT} does not change. Otherwise
\equ{coordT} should be considered as the `$T$--component' of a 4--vector.

\medskip

The relation between $SL(2,\Cset)$ and Lorentz group \equ{weyltr} is valid
not only for null vectors. Any vector is a sum of two null vectors and
\par\centerline{$\mathbf{A(V+U)A^* = AVA^* + AUA^*}$.}

The qubit is described by two-component complex vector or {\em Weyl spinor}.
It corresponds to massless particle with spin 1/2. Such particle always moves
with the speed of light. The equations \equ{coord4} show a correspondence
between such spinor and 4D null vector (\fig{isotr}). This vector can be also
rewritten by using Pauli matrices:
\begin{equation}
\matrix{
 \sigma_x = \left(\matrix{0&1\cr 1&0\cr}\right)\!,\ \
 \sigma_y = \left(\matrix{0&-i\cr i&0\cr}\right)\!,\ \
 \sigma_z = \left(\matrix{1&0\cr 0&-1\cr}\right)\cr\cr
 \mathbf V = T \mbf{1} + X \sigma_x + Y \sigma_y + Z \sigma_z, \hfill\cr
 V_i = \frac12 Tr({\sigma_i \mathbf V}) = Tr(\sigma_i \psi \psi^*) =
 \psi^* \sigma_i \psi; \hfill\cr
 \mbf{\sigma} = \{\sigma_x,\sigma_y,\sigma_z\}:\ \
(T,\{X,Y,Z\})=(\psi^*\psi,\psi^*\mbf{\sigma}\psi) \hfill\cr
}
\label{eq:sigm}
\end{equation}

\subsection{Massive particle}

Massive charged particle with spin 1/2 like an electron is described by
two Weyl spinors and has four complex components:

\begin{equation}
\psi = \left(\matrix{\varphi_R \cr \varphi_L}\right)
\  \varphi_R, \varphi_L \in \Cset^2;
\quad \psi = \left(\matrix{\psi_0 \cr \psi_1 \cr \psi_2 \cr \psi_3}\right)
\label{eq:bispinor}
\end{equation}

It is possible to consider such massive particle as two qubits:
\begin{equation}
\psi = c_{00}|00\rangle + c_{01}|01\rangle +
       c_{10}|10\rangle + c_{11}|11\rangle
\label{q2bit}
\end{equation}
The first index is similar to $|\!\uparrow\rangle$ and $|\!\downarrow\rangle$
for each $\varphi_R, \varphi_L$. The other one corresponds to discrete
coordinate transformation like {\em spatial reflection\/}:
\mbox{$P:(t,\vec{x})\to(t,-\vec{x})$}.

\begin{figure}[!htb]
\centering
 \begin{picture}(180,150)
 \small
 \put(0,0){\epsfig{file=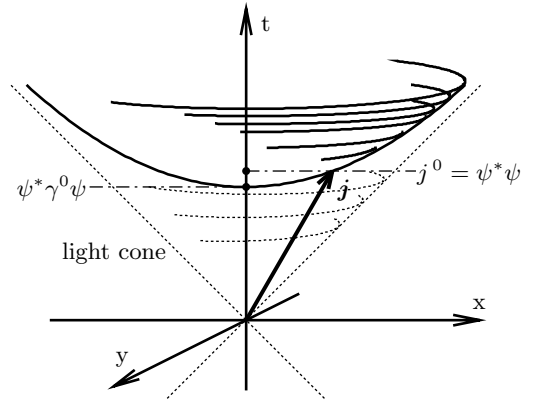}}
 \put(59,60){\makebox(0,0)[tr]{light cone}}
 \put(95,140){t}
 \put(175,35){x}
 \put(40,15){y}
 \put(124,84){\makebox(0,0)[tl]{\boldmath $j$}}
 \put(154,86.75){\makebox(0,0)[l]{$j\,^0 = \psi^*\psi$}}
 \put(29,80.75){\makebox(0,0)[r]{$\psi^*\gamma^0\psi$}}
 \end{picture}
\caption{Massive particle}\label{fig:curr}
\end{figure}

It is also possible to build a vector by using the 4D spinor and $4 \times 4$
Dirac matrices $\gamma^{\mu}$. It is 4D vector of {\em current\/} \fig{curr} :
\begin{equation}
 j^{\mu} = \psi^* \gamma^0 \gamma^{\mu} \psi \\
\label{eq:ecurr}
\end{equation}
\begin{equation}
\gamma^0 = \left(\matrix{0&1\cr 1&0\cr}\right)\! ,\ \
\mbf{\gamma} = \left(\matrix{0&\mbf{-\sigma}\cr \mbf{\sigma}&0\cr}\right).
\label{eq:dirmtar}
\end{equation}
with always positive:
\begin{equation}
  j^0 = \psi^* \psi = {\textstyle \sum_i}|\psi_i|^2 = ||\varphi_R||^2+||\varphi_L||^2
\label{eq:ej0}
\end{equation}
but $j^0$ is not Lorentz invariant. The Lorentz invariant scalar is
\begin{equation}
  \psi^* \gamma^0 \psi = \varphi_R^*\varphi_L^{} + \varphi_L^*\varphi_R^{}
\label{eq:ejm}
\end{equation}

\subsection{Representations of Lorentz group}

We have used very simple construction of a qubit, but
any other constructions also have limitations because a representation
of Lorentz group cannot satisfy contemporary two following conditions:
\footnote{A ``{\em no-go}'' result for bounded quantum networks.} 
 \begin{itemize}
  \item The representation is finite dimensional.
  \item The representation is unitary in a {\em definite\/} norm.
 \end{itemize}
It can be considered as some mathematical reasons for:
 \begin{itemize}
  \item Using of {\em quantum field theory\/} (QFT) instead of systems with
        finite number of states.
  \item Necessity of a consideration of different kinds of interacting
        quantum fields.
 \end{itemize}

The relativistic physics have both these properties. We can consider
{\em Quantum Electrodynamics\/} (QED) as an example.

It is not quite compatible with such properties of usual model of quantum
computation as fixed size of registers and gates, one kind of {\em qubits},
etc.\@.

\section{Quantum field theory and computations}\label{qft}

In articles about quantum computers
Feynman \cite{feynman:simul,feynman:comp} has used one of usual tools of a QFT
--- {\em annihilation\/} and {\em creation\/} operators $a$ and $a^{*}$:

\begin{equation}
\begin{array}{c}
 a   = \left(\matrix{0&0\cr
                     1&0\cr}\right), \quad
 a^* = \left(\matrix{0&1\cr
                     0&0\cr}\right), \\
 N \equiv a^* a
     = \left(\matrix{1&0\cr
                     0&0\cr}\right)
\end{array}
\label{eq:pauli}
\end{equation}
with Fermi relation for the anticommutator:
\begin{equation}
  \{a^*,a\}_+ \equiv a^* a + a \, a^* = 1
\label{eq:anticomm}
\end{equation}

These operators are used for describing of usual quantum gate in
\cite{feynman:comp}, but this approach has more wide scope. This method has
a resemblance with secondary quantization in a QFT.

\subsection{Secondary quantization}

In a QFT wave functions are operators \cite{bs:itqf}.
Let us consider photons as an example:

\begin{equation}
    \hat{\psi}_p = c_p e^{-i p x} + c^{*} _p e^{i p x}
\label{eq:photons}
\end{equation}

There $c_p$ and $c^{*} _p$ are operators of annihilation and creation
of the particle with 4--momentum {\em p\/} and so $\hat{\psi}$ is an operator.
There is Bose relation for the commutator:
\begin{equation}
  [c^*,c]_- \equiv c^* c - c \, c^* = 1
\label{eq:comm}
\end{equation}

\subsection{States and operators}

The operators $c_p$ and $c^{*} _p$ act in some auxiliary Hilbert space
and functions like \equ{photons} have more direct physical
meaning than states in this space. The quantum field of electrons is
described by some expression similar to \equ{photons}\footnote{
The main difference is commutational relations \equ{anticomm} for electrons
and \equ{comm} for photons.}.

The matrices \equ{pauli} are used for presentation of quantum gates in
\cite{feynman:comp}, but it should be mentioned that in relativistic physics
there is no sharp division between q-gates and q-states due to formulae like
\equ{photons}.

This property of a QFT has some analogy with functional style of programming in
modern computer science \cite{backus:fp}. In both cases there is no essential
difference between data (states) and functions (operators). A function can be
used as data for some other function.

\medskip

\begin{figure}[!htb]
\centering
\small
\unitlength=1mm
\special{em:linewidth 0.4pt}
\linethickness{0.4pt}
\begin{picture}(30.00,15.00)
\put(2.50,2.50){\framebox(5.00,10.00)[cc]{}}
\put(2.50,7.50){\line(1,0){5.00}}
\put(12.50,10.00){\vector(-1,0){5.00}}
\put(12.50,5.00){\vector(-1,0){5.00}}
\put(12.50,2.50){\framebox(7.50,10.00)[cc]{}}
\put(23.50,10.00){\vector(-1,0){3.50}}
\put(23.50,5.00){\vector(-1,0){3.50}}
\put(23.50,2.50){\framebox(5.00,10.00)[cc]{}}
\put(23.50,7.50){\line(1,0){5.00}}
\put(5.00,13.50){\makebox(0,0)[cb]{$\psi'$}}
\put(10.00,13.50){\makebox(0,0)[cb]{$=$}}
\put(16.50,13.50){\makebox(0,0)[cb]{$U$}}
\put(26.00,13.50){\makebox(0,0)[cb]{$\psi$}}
\end{picture}
\caption{Nonrelativistic gate}\label{fig:qgate}
\end{figure}
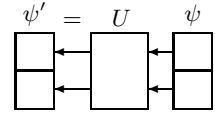

For example, let us consider an electron as the model of a qubit. In
nonrelativistic quantum theory of computation a q-gate can change state of
the qubits $\psi' = U \psi$ (\fig{qgate}). Here $\psi,\psi'$ are wave vectors
of quantum system (`qubits') and $U$ is an operator of the gate.

\begin{figure}[!htb]
\centering
 \begin{picture}(140,60)
 \put(0,0){\epsfig{file=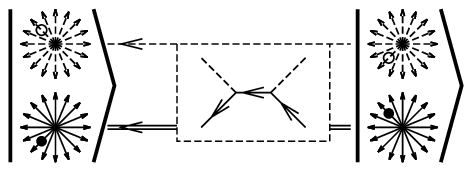}}
 \small
 \put(77.00,50.00){\makebox(0,0)[cb]{$\mathcal{U}_{\hat{\psi},\hat{A}}$}}
 \put(20.00,50.00){\makebox(0,0)[cb]{$\Psi'$}}
 \put(45.00,50.00){\makebox(0,0)[cb]{$=$}}
 \put(120.00,50.00){\makebox(0,0)[cb]{$\Psi$}}
 \put(11.00,44.50){\makebox(0,0)[cc]{$\gamma$}}
 \put(11.00,20.50){\makebox(0,0)[cc]{\boldmath$e$}}
 \end{picture}
\caption{Relativistic gate}\label{fig:rqgate}
\end{figure}

The gate can be built as some electro-magnetic device. From point of view of
QED it is described as an interaction of two quantum fields and we should not
split the processes on q-gates and qubits. The usual formula of secondary
quantization is $\Psi' = \mathcal{U}_{\hat{\psi},\hat{A}} \Psi$ (\fig{rqgate}). Here
$\Psi,\Psi'$ describe {\em occupation numbers\/}, and $\hat{\psi}$ is wave
operator for electron (positron), and $\hat{\mathbf A}$ for photons. The
wave operators for particle are included in $\mathcal{U}$ and can form many
{\em nonlinear\/} expressions. They correspond to Feynman diagrams. Such
description is linear in respect of $\Psi,\Psi'$, but not on $\hat{\psi},\hat{A}$.

\subsection{Algebraic and matrix notation}

The relations \equ{anticomm} and \equ{comm} describe one particle. If
we have a few particles then the full set of relations is:

\begin{equation}
\begin{array}{l}
  \{a_k, a_{k'} \}_+ = \{a_k^*, a_{k'}^* \}_+ = 0 \cr
  \{a_k, a^*_{k'} \}_+ = \delta _{k k'}
\end{array}
\label{eq:anticommut}
\end{equation}
for particles like electrons (Fermi statistic, half-integer spin) and
\begin{equation}
\begin{array}{l}
  [c_k, c_{k'}]_- = [c_k^*, c_{k'}^*]_- = 0 \cr
  [c_k, c^*_{k'}]_- = \delta _{k k'}
\end{array}
\label{eq:commut}
\end{equation}
for particles like photons (Bose statistic, integer spin).

\medskip

The equations \equ{pauli}, \equ{anticomm} show representation of
operators with Fermi relations for one particle. The matrix representations
of \equ{anticommut} for many particles are more complicated.

The relations for Bose particles \equ{comm}, \equ{commut} are impossible
to express by using finite-dimensional matrices because for any two matrices
${\bf A}, {\bf B}$: \begin{equation}
  Trace({\bf A} {\bf B} - {\bf B} {\bf A}) = 0 \quad \Longrightarrow
  \quad [{\bf A},{\bf B}]_-   \neq {\bf 1}
\label{eq:ab}
\end{equation}

Due to such properties of algebras of commutators the presentation by using
formal expressions with operators of annihilation and creation
\cite{feynman:comp} instead of matrices can be more convenient in quantum
theory of computation from the point of view of relativistic physics.

\section{Conclusion}\label{concl}

In nonrelativistic quantum theory of computation it was necessary only to
point number of states $2^n$ for description of q$^n$bit. In relativistic
theory there are many special cases. The charged and neutral, massive and
massless particles {\em etc.\/} should be described differently.

\section*{Acknowlegments}
The author is grateful to organizers of the PhysComp96, especially to Tommaso
Toffoly, and to Physics Department of Boston University for support and hospitality.


\end{document}